\documentclass[showpacs,amsmath,amssymb,aps,showkeys,floatfix,prd,a4paper]{revtex4}
\usepackage[dvips]{graphicx}
\usepackage{dcolumn}
\usepackage{bm} %bold math
\usepackage{epsfig}
\usepackage{amsfonts}
\usepackage{amssymb,amscd}
\usepackage{subfigure}
\usepackage{xcolor}
\usepackage{verbatim}

\def\pom{{I\!\!P}}

\newcommand{\Pelotas}{High and Medium Energy Group, Instituto de F\'isica e Matem\'atica,
             Universidade Federal de Pelotas\\
             Caixa Postal 354,  96010-900, Pelotas, RS, Brazil.}

\begin{document}

\title{Exclusive and diffractive quarkonium -- pair production  at the LHC and FCC}
\author{ V. P. Gon\c calves and R. Palota da Silva}
\affiliation{ \Pelotas}

\begin{abstract}
The production of a quarkonium -- pair in exclusive and diffractive processes in $pp$ collisions at the LHC and FCC energies is investigated. We consider the $J/\Psi J/\Psi$ and $\Upsilon \Upsilon$ production in these processes and present predictions for  the transverse momentum and rapidity distributions  considering the kinematical ranges expected to be covered by central and forward detectors. Results for the cross sections are also presented. Our results indicate that the double $J/\Psi$ production is dominated by the exclusive process, while the double  $\Upsilon$ production receive a large contribution of the diffractive process.  The impact of the modelling of the gap survival factor on our predictions is discussed. 
\end{abstract}
%\pacs{12.38.-t; 13.60.Le; 13.60.Hb}


\maketitle

%\section{Introduction}

In the last years the production of a quarkonium -- pair ($J/\Psi J/\Psi$ and $\Upsilon \Upsilon$) have been the subject of a large number of studies 
\cite{Kart,Hump,Vogt,Qiao,Li,Qiao_jpg,Ko,Bere,trunin,lans4,Li_relativistic,lans2,lans3,Sun,lans1,
Kom_double,Baranov1,Baranov2,david_nuc,rafal_double,kulesza,Bere2}. Such analysis were strongly motivated by the experimental data from the D0 \cite{d0} Collaboration at Tevatron and the ATLAS, CMS and LHCb Collaborations at the LHC \cite{atlas,cms,lhcb}, which have indicated that the contribution of the  double partonic scattering is not negligible in inclusive proton -- proton collisions, where both colliding protons dissociate in the interaction. In addition, the LHCb Collaboration have also released the first data for the exclusive double $J/\Psi$ production \cite{lhcb_double}, where both colliding protons remain intact and the final state is characterized by two rapidity gaps, i.e. empty regions  in pseudo-rapidity that separate the intact very forward protons from the $J/\Psi J/\Psi$ state. The study of the exclusive reactions is expected to improve our understanding about the  interplay between the small- and large-distance regimes of Quantum Chromodynamics (QCD) \cite{review_forward}. As pointed out in Refs. \cite{Mariotto,Nos_prd}, a quarkonium -- pair can also be produced in double diffractive reactions, which also are characterized by two rapidity gaps and two intact protons in the final state, but the quarkonium pair is generated by the interaction between gluons  of the Pomeron ($\pom$), which is a color singlet object inside the proton.
The results presented in Ref. \cite{Nos_prd} indicated that the contribution of the double diffractive reaction for the double $J/\Psi$ production are similar to those derived in Ref. \cite{khoze} considering the exclusive production.  Such result motivates the analysis to be performed in this paper, where we will present a comparison between the predictions for the double quarkonium production in diffractive and exclusive reactions. In particular, we will update the predictions derived in Ref. \cite{Nos_prd} considering a more recent parameterization for the diffractive gluon distribution, which was obtained in Ref. \cite{Guzey} by fitting the latest HERA data for diffractive $ep$ reactions. Moreover, we will present, for the first time, our predictions for the diffractive double quarkonium production at the energies of the Future Circular Collider (FCC) \cite{fcc}. In addition, we will extend the formalism present in Ref. \cite{khoze} for the double $\Upsilon$ production and estimate, for the first time, the prediction of this final state in exclusive reactions at the LHC and FCC. Our goal is to determine the kinematical range of dominance of these two reactions channels in order to be able to use  future experimental data to constrain the underlying assumptions present in the description of diffractive and exclusive processes.

\begin{figure}[t]
\begin{tabular}{ccc}
{
\includegraphics[scale =.7]{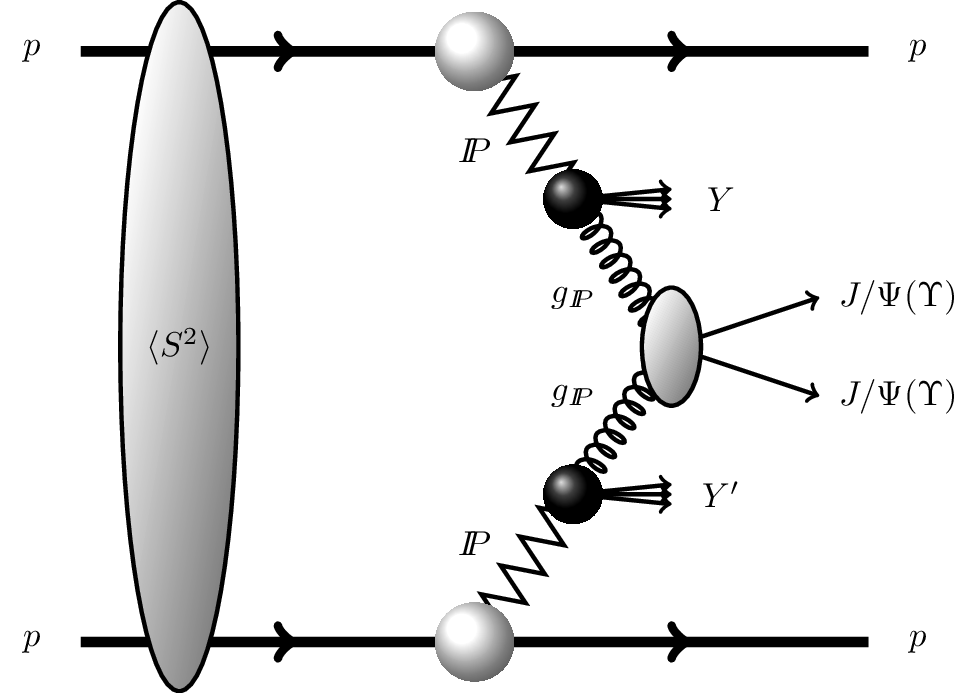}
} & \,\,\, &
{
\includegraphics[scale =.7]{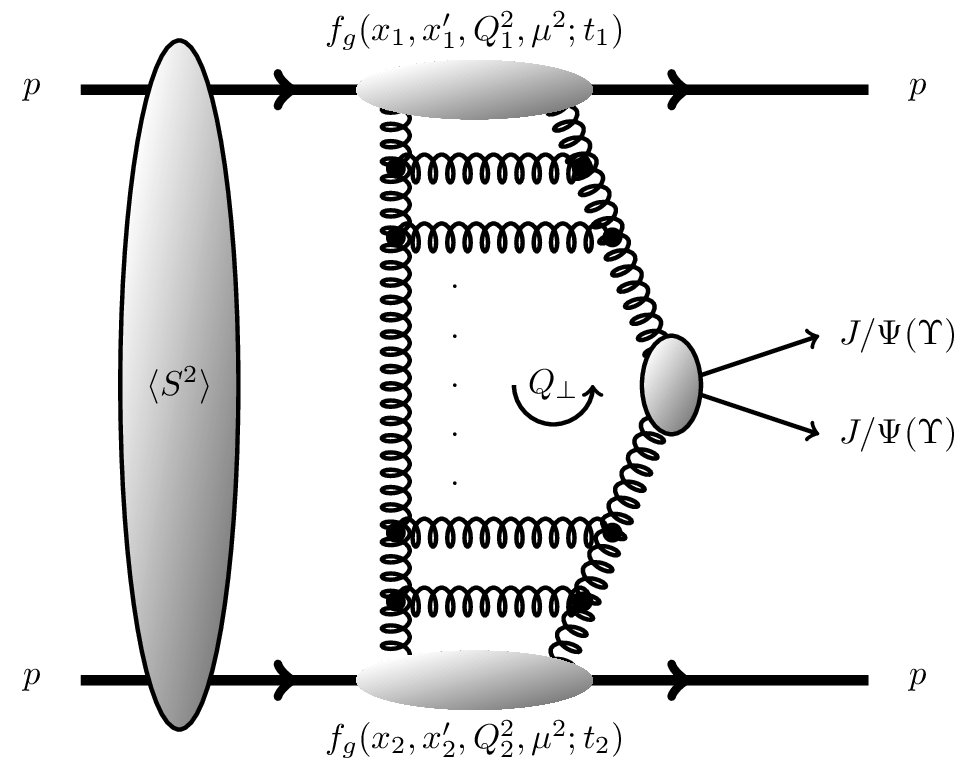}
}
\end{tabular}
\caption{Typical diagrams for the quarkonium -- pair production in  double diffractive (left panel) and exclusive (right panel)  processes. The blob denoted by  $\langle S^2\rangle$ represents the gap survival factor associated to absorptive effects (See text for details). } 
\label{Fig:Diagramas}
\end{figure}

The typical diagrams that characterize the quarkonium -- pair production in double diffractive and exclusive processes are represented in  Fig. \ref{Fig:Diagramas}. 
Differently from the exclusive case, where only the quarkonium -- pair will be present in the final state, we have in the double diffractive process the presence of  additional particles in the final state, associated to the remnants of the Pomeron, which dissociates in the interaction. The presence of these particles can be used, in principle, to discriminate the diffractive from the exclusive double quarkonium production. However,  due to the large pile-up of events in each bunching crossing expected to be present in the future runs of LHC and FCC, it is not clear if the separation of the  diffractive and exclusive events will be possible by measuring the rapidity gaps and counting the number of tracks in the final state. As a consequence, events characterized by two rapidity gaps probably only will  be separated  by tagging the intact hadrons in the final state using forward detectors, as e.g. the CT-PPS and AFP forward proton spectrometers, associated with the ATLAS and CMS central detectors \cite{detectors}. As exclusive and double diffractive events are characterized by intact hadrons, it is fundamental to know what is the relative contribution of each of these processes for the quarkonium -- pair production at central and forward rapidities as well as for LHC and FCC energies.
In what follows we will present a brief review of the formalism proposed to describe these reactions, which is based on the Resolved Pomeron model \cite{IS} in the diffractive case and in the Durham model \cite{kmr_prosp} in the exclusive case.
The cross section for the double diffractive cross section can be estimated using the nonrelativistic QCD (NRQCD) factorization formalism \cite{nrqcd} for the quarkonium production   and  is given by \cite{Nos_prd}:
\begin{eqnarray}
d\sigma (pp\rightarrow p + {\cal{Q}} {\cal{Q}} + p) = \sum_n g^D_{p} (x_1,\mu^2)\, g^D_{p} (x_2,\mu^2)   \cdot d\hat{\sigma}[gg\to Q\bar{Q}_n+Q\bar{Q}_n] \cdot \langle {\cal{O}}^{{\cal{Q}}}_{n} \rangle \langle {\cal{O}}^{{\cal{Q}}}_{n} \rangle \,\,,
\label{csdd}
\end{eqnarray}
where ${\cal{Q}}$ represents the quarkonium state, $g^D$ is the diffractive gluon distribution probed at the scale $\mu^2$, which we assume to be equal to $\mu^2 = M^2 + p_T^2$, and  
$d\hat{\sigma}$  is the differential cross section for the $gg\to Q\bar{Q}_n+Q\bar{Q}_n$ subprocess, which is perturbatively calculated considering the production of the heavy quark pair $Q\bar{Q}$ in an intermediate Fock state $n$, which does not have to be color neutral. Moreover, the  quantities $\langle {\cal{O}}^{{\cal{Q}}}_n\rangle$
are nonperturbative long distance matrix elements, which describe the transition of the intermediate $Q\bar{Q}$ in the physical state ${{\cal{Q}}}$ via soft gluon radiation and are determined from experimental data. As in Ref. \cite{Nos_prd}, we will estimate $d\hat{\sigma}$
taking into account of the 31 diagrams that contribute for the color singlet channel, as well as the 72 diagrams for the color-octet channel. Moreover, we will assume the following values for the 
 color -- octet large-distance matrix elements:  $\langle O_8^{J/\psi}({}^3S_1) \rangle=3.9\times 10^{-3}$ GeV$^3$  \cite{Braaten2000} and $\langle O_8^{\Upsilon}({}^3S_1) \rangle=1.5\times 10^{-1}$ GeV$^3$ \cite{kramer}. In the case of the color -- singlet channel, it is possible to express the matrix elements in terms of  the square of the radial wave function of the quarkonium ${\cal{Q}}$ at the origin ($|R_{{\cal{Q}}}(0)|^2$) which is related to the leptonic decay rate 
$\Gamma({\cal{Q}}\to e^+e^-)$ \cite{Eichten}. Therefore, such quantity can be determined e.g. using the recent PDG data \cite{pdg}. 
Finally, in order to calculate the double diffractive cross section we should to assume a model for the diffractive gluon distribution. As in our previous studies \cite{Mariotto,Nos_prd}, we will describe this quantity using the Resolved Pomeron model \cite{IS}, which implies that  
$g^D_{p} (x,Q^2)$ is defined as a convolution of the \,{Pomeron} flux emitted by the proton, $f^{p}_{\pom}(x_{\pom})$, and the gluon distribution in the \,{Pomeron}, $g_{\pom}(\beta, Q^2)$,  with  $\beta$ being the momentum fraction carried by the partons inside the \,{Pomeron}.
Such quantities can be constrained using the experimental data from diffractive deep inelastic scattering (DDIS) at HERA. Differently from Ref. \cite{Nos_prd}, where we have used the 
 fit B  obtained by the H1 Collaboration at DESY-HERA  \cite{H1diff} several years ago, in what follows we will consider the more recent parameterization obtained in Ref. \cite{Guzey} using the high -- precision data from H1/ZEUS combined inclusive diffractive cross sections measurements.
We will use the fit A proposed in Ref. \cite{Guzey}, but we have verified that the results obtained using the fit B are very similar.

The central exclusive processes are usually described using the Durham model \cite{kmr_prosp}, proposed many years ago and extensively discussed in the literature (For a review see, e.g. Ref. \cite{review_kmr}). 
In this  model, the cross section  for the central exclusive  production of a quarkonium -- pair can be expressed in terms of the 
skewed unintegrated gluon distributions $f_g$ and the sub--amplitude for the $gg \rightarrow 
{\cal{Q}} {\cal{Q}}$ process \cite{khoze}. At leading logarithmic approximation, it is possible to express $f_g(x,x^{\prime},Q_t^2, \mu^2)$ in terms of the conventional integral gluon density $g(x)$ and the Sudakov factor $T$, which ensures that the active gluons that participate of the hard process do not radiate in the evolution from $Q_t$ up to the hard scale $\mu = m_{\perp} \equiv  \sqrt{ M_{{\cal{Q}}}^2 + p_{{\cal{Q}},\perp}^2}$. The amplitude for the $gg \rightarrow 
{\cal{Q}} {\cal{Q}}$ process can be estimated using  the hard exclusive formalism proposed in Refs. \cite{brodsky_lepage,chernyak} and  considering the non -- relativistic approximation. 
The results presented in Ref. \cite{khoze} demonstrated that the exclusive reaction is only sensitive to the color -- singlet component of the meson wave function, do not receiving color -- octet contributions.
The final expressions for the double $J/\Psi$ production have been included in the publicly available SuperChic Monte Carlo (MC) \cite{superchic}. In order to also estimate the double $\Upsilon$ production, we have modified the SuperChic and included this final state, which allow us to perform a full MC simulation of quarkonium -- pair production in central exclusive processes. As in Ref. \cite{khoze} we have fixed the value of the $\Upsilon$ wave function at the origin to its leptonic width. Moreover, in our calculations we have considered that the conventional gluon distribution is given by the  MMHT2014 parameterization \cite{mmht}.

{
One important open question in the description of central exclusive and diffractive interactions in $pp$ collisions  is the treatment of the  soft interactions that are expected to lead to extra production of particles, which will  destroy the rapidity gaps in the final state and modify the associated cross sections \cite{bjorken}.  The experimental results from Tevatron and LHC for these processes have demonstrated that these additional absorption effects cannot be neglected. Theoretically, the soft 
 rescattering corrections associated to reinteractions (often
referred to as multiple scatterings) between spectator partons of the colliding protons imply 
the violation of the QCD hard scattering factorization theorem for diffraction \cite{collinsfac}. Such corrections  modify the Resolved Pomeron and Durham predictions for the diffractive and central exclusive processes. The modelling of the
soft multiple scattering in diffrative $pp$ collisions have been the subject of several studies during the last years. For example, in Refs. \cite{torbcris,qgsjetii,herwig} the authors have proposed to treat these effects using a general purpose Monte Carlo. However, such approaches are still strongly
dependent on the treatment of the multiple interactions, the assumptions for the color flow along the rapidity gap as well as the modelling of possible proton excitations.
In the case of the exclusive processes, the Durham group have proposed an approach to treat the absorptive corrections associated to the additional soft proton -- proton interactions (denoted eikonal factor $S^2_{\mbox{eik}}$), which are independent of the hard processes, as well the rescatterings of the protons with the intermediate partons that are described by the so--called enhanced factor $S^2_{\mbox{enh}}$. As discussed in Ref. \cite{khoze}, the magnitude of the enhanced factor is still uncertain, but it is expected to generate a weaker suppression in comparison to that associated to the eikonal survival factor. In the case of the predictions for the double $J/\Psi$ production in exclusive processes presented in Ref. \cite{khoze}, the enhanced corrections were not included in the calculations. In what follows we also will assume this  approximation. In Table \ref{Tab:1} we present the predictions for the total cross sections for the quarkonium -- pair production in $pp$ collisions at the LHC and FCC considering the four different models for  the  eikonal factor $S^2_{\mbox{eik}}$ present in the SuperChic MC. One have that the distinct treatments of  $S^2_{\mbox{eik}}$ implies that the predictions can differ by a factor $\approx 3$.
Another possible approach, largely used in the literature  (See e.g.  Refs. 
\cite{nosbottom,MMM1,antoni,antoni2,cristiano,kohara_marquet,nos_ze,nos_Dijet,nos_dimuons,nos_last}),  
is based on the assumption that the hard process occurs on a short enough timescale such that the physics that generate the additional particles can be factorized and accounted by an overall factor, denoted gap survival factor $\langle S^2\rangle$, multiplying the cross section \cite{kkmr}. 
In general the values of $\langle S^2\rangle$ depend on the energy, being typically of order 0.01 -- 0.05  for LHC energies. In particular, for the   quarkonium -- pair production in double diffractive interactions in $pp$ collisions at $\sqrt{s} = 14$ (100) TeV it is expected to be 0.02 (0.01), i.e. the absorptive corrections are expected to suppress the cross section by a factor 50 (100).
For comparison, in Table \ref{Tab:1} we also present the results derived  multiplying by $\langle S^2\rangle$ the SuperChic predictions calculated without the inclusion of the absorptive corrections.
For LHC energy and the double $J/\Psi$ production, the resulting predictions are similar to those derived with $S^2_{\mbox{eik}}$. For FCC energy, the prediction is slightly larger. In contrast, for the double $\Upsilon$ production, we predict larger cross sections assuming the overall factor $\langle S^2\rangle$
instead of $S^2_{\mbox{eik}}$. As the modelling, magnitude and universality  of  absortive corrections  {are still} a theme of intense debate \cite{review_kmr,durham,telaviv}, in what follows we will assume that the absorptive corrections for the double diffractive and central exclusive processes can be modelled by the same factor  $\langle S^2\rangle$. Surely such assumption can and must be improved in the future. However, considering the current large theoretical uncertainty in the treatment of the  soft interactions, we believe that such simplistic approach allow us, at least,  to understand what are the main differences  between the diffractive and exclusive quarkonium -- pair production associated to the distinct approaches for the hard process.

\begin{table}[t]
	\centering
	\begin{tabular}{||l| r| r||}
		\hline
		Absorptive factor & $\sigma_{2J/\Psi}$ [pb] & $\sigma_{2\Upsilon}$ [pb] \\
		\hline
		\hline
		$\langle S^2_{\mbox{eik}} \rangle$ -- Model 1 & 13.8 (132.1) & 9.3$\times 10^{-5}$ (8.1$\times 10^{-4}$) \\
		\hline
		$\langle S^2_{\mbox{eik}} \rangle$ -- Model 2 & 48.1 (352.5) & 2.4$\times 10^{-4}$ (2.2$\times 10^{-3}$) \\
		\hline
		$\langle S^2_{\mbox{eik}} \rangle$ -- Model 3 & 37.7 (250.1) & 1.6$\times 10^{-4}$ (1.7$\times 10^{-3}$) \\
		\hline
		$\langle S^2_{\mbox{eik}} \rangle$ -- Model 4 & 20.6 (196.7) & 1.2$\times 10^{-4}$ (1.2$\times 10^{-3}$)\\
		\hline
		$\langle S^2 \rangle$ = 0.02 (0.01) & 39.3 (439.2) & 1.6$\times 10^{-3}$ (2.4$\times 10^{-2}$) \\
		\hline
	\end{tabular}
	\caption{Total cross sections for the quarkonium -- pair production in central exclusive processes considering $pp$ collisions at $\sqrt{s} = 14$ TeV and different models for the absorptive factor. Values in parentheses are for FCC energies ($\sqrt{s} = 100$ TeV).}
	\label{Tab:1}
\end{table}

\begin{figure}
\begin{tabular}{c c}
{
\includegraphics[scale = .4]{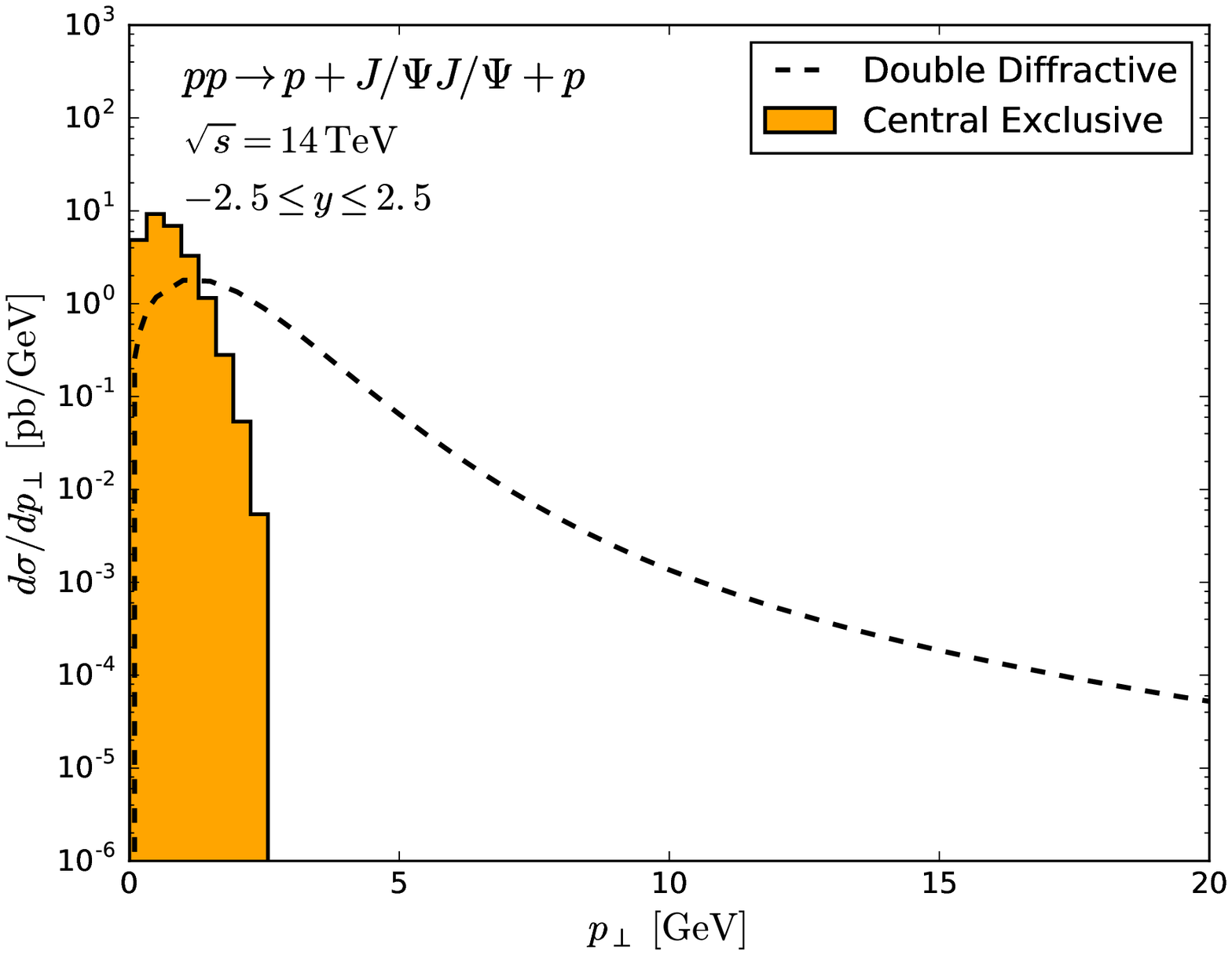}
}
{
\includegraphics[scale = .4]{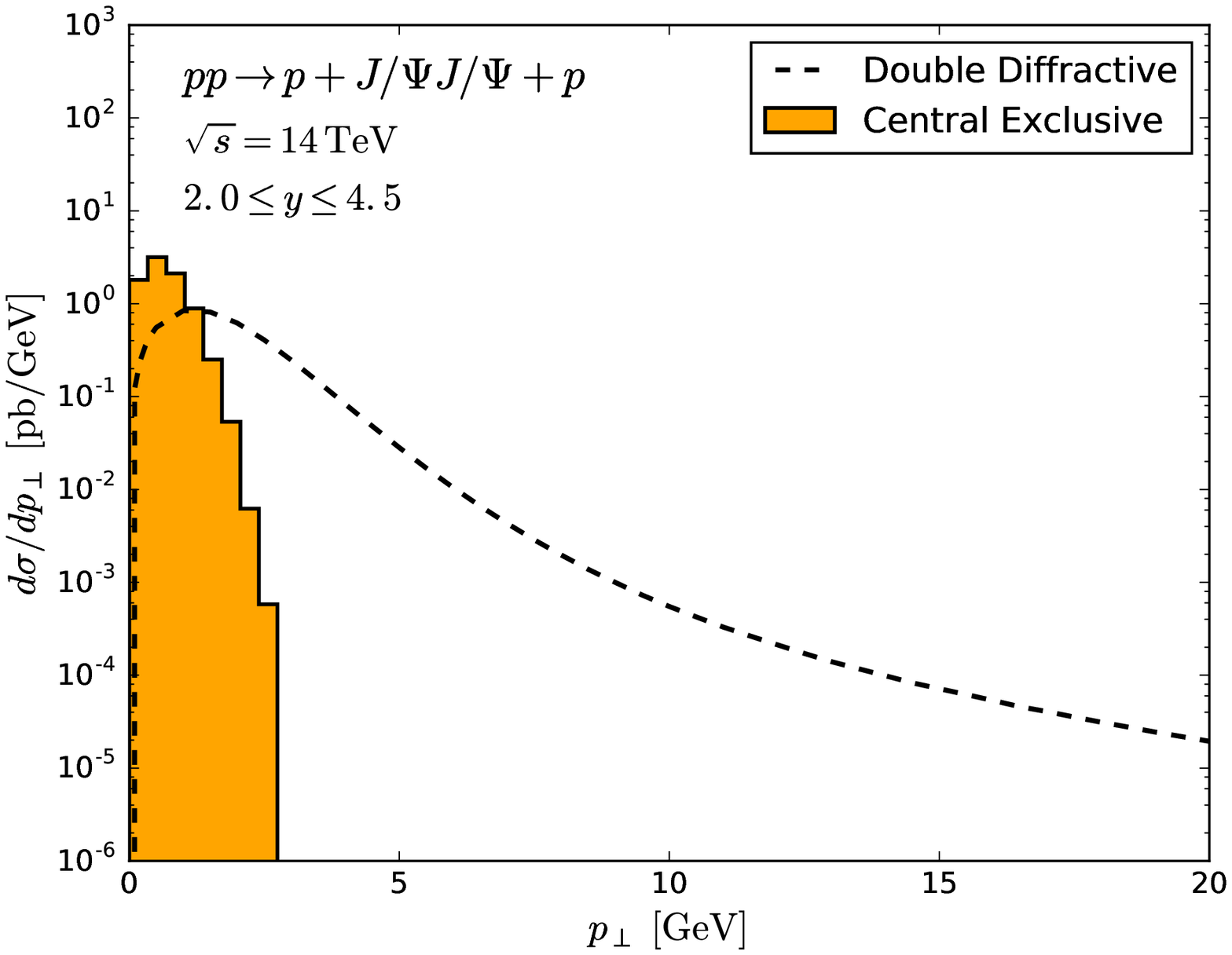}
} \\
{
\includegraphics[scale = .4]{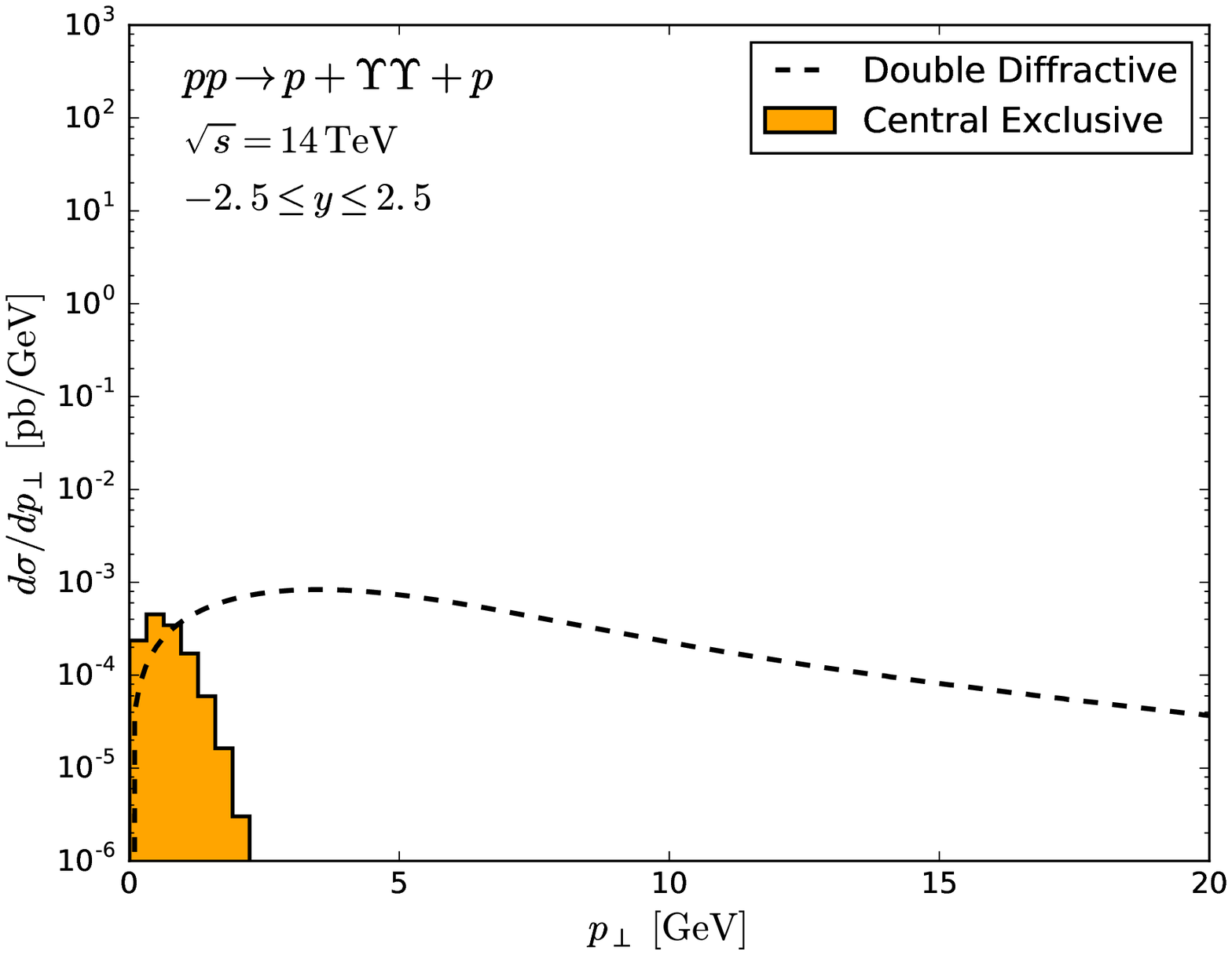}
}
{
\includegraphics[scale = .4]{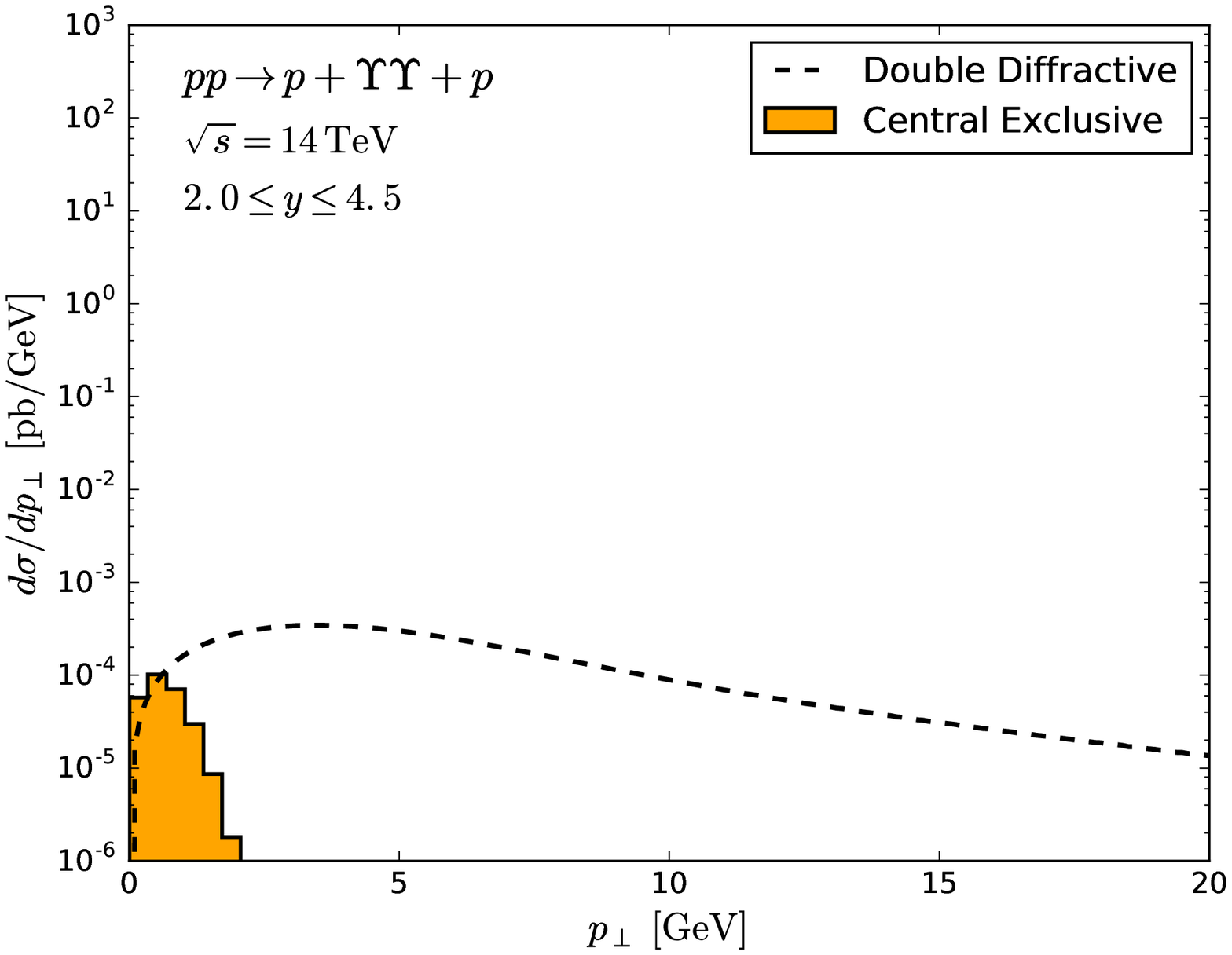}
}

\end{tabular}
\caption{Transverse momentum distributions for the  diffractive and central exclusive $J/\Psi J/\Psi$ (upper panels) and  $\Upsilon \Upsilon$ (lower panels) production in $pp$ collisions at $\sqrt{s} = 14 $ TeV considering the rapidity ranges covered by a central (left panels) and a forward (right panels) detector. } 
\label{Fig:TransMom}
\end{figure}

\begin{figure}
\begin{tabular}{c c}
{
\includegraphics[scale = .4]{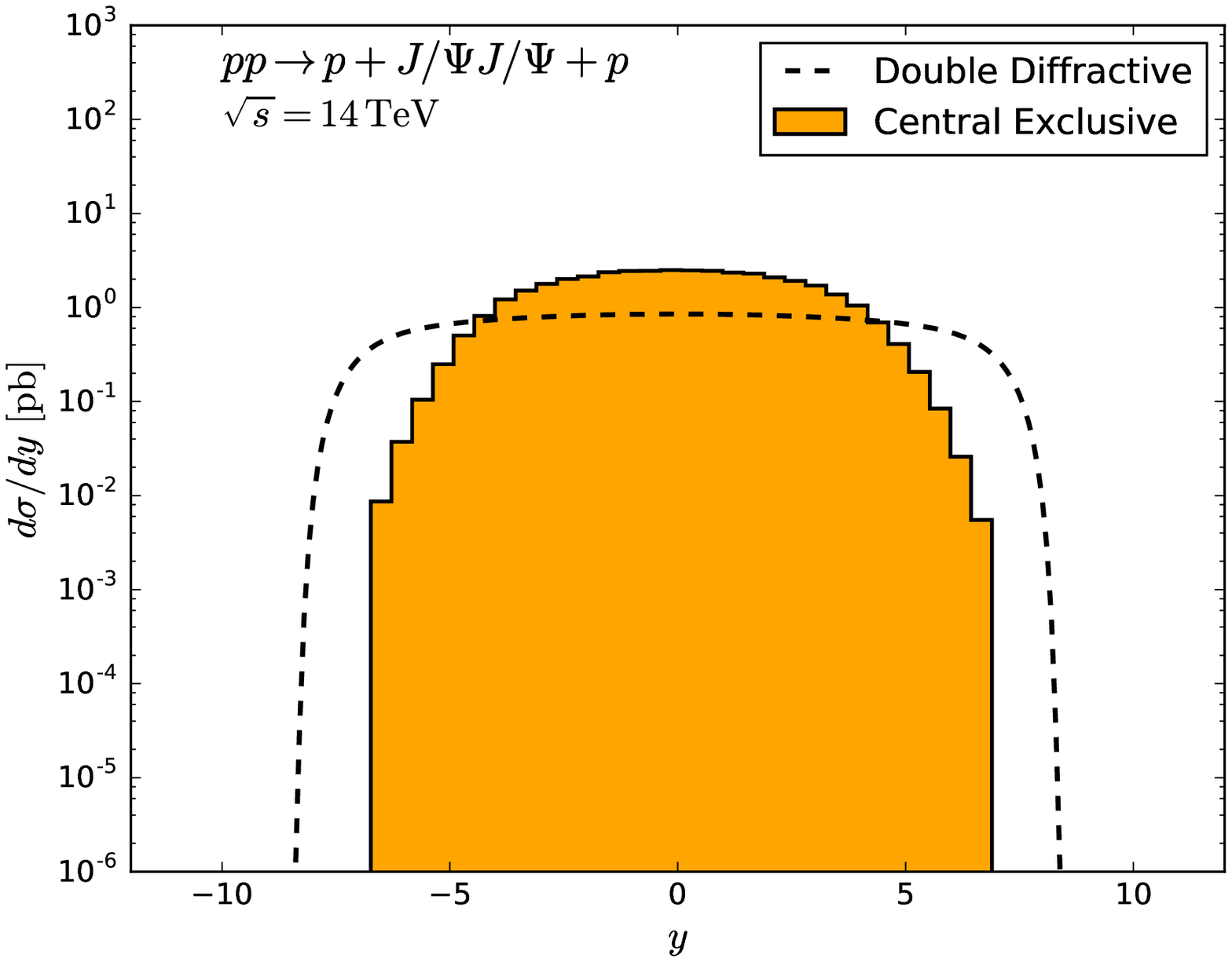}
}
{
\includegraphics[scale = .4]{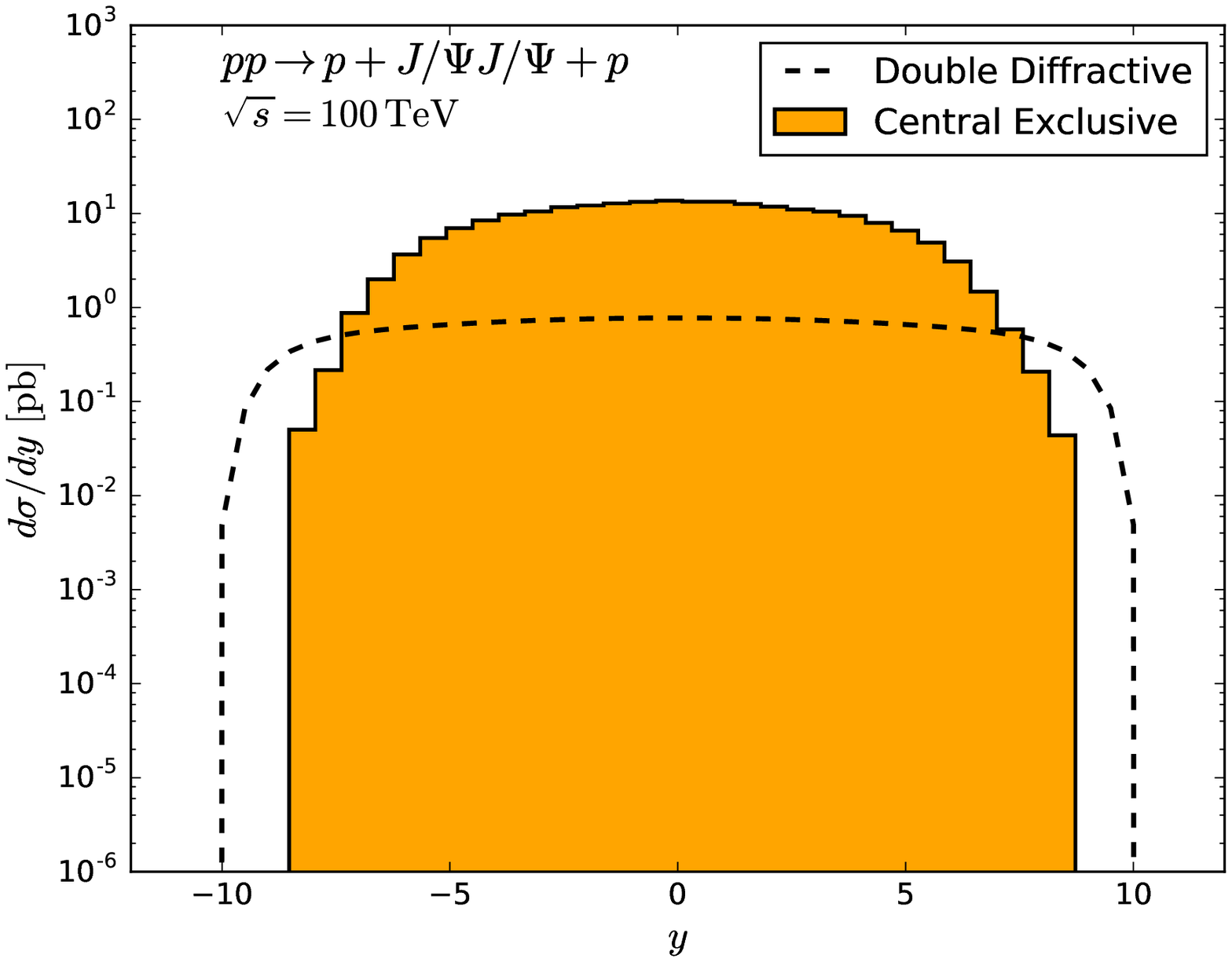}
} \\
{
\includegraphics[scale = .4]{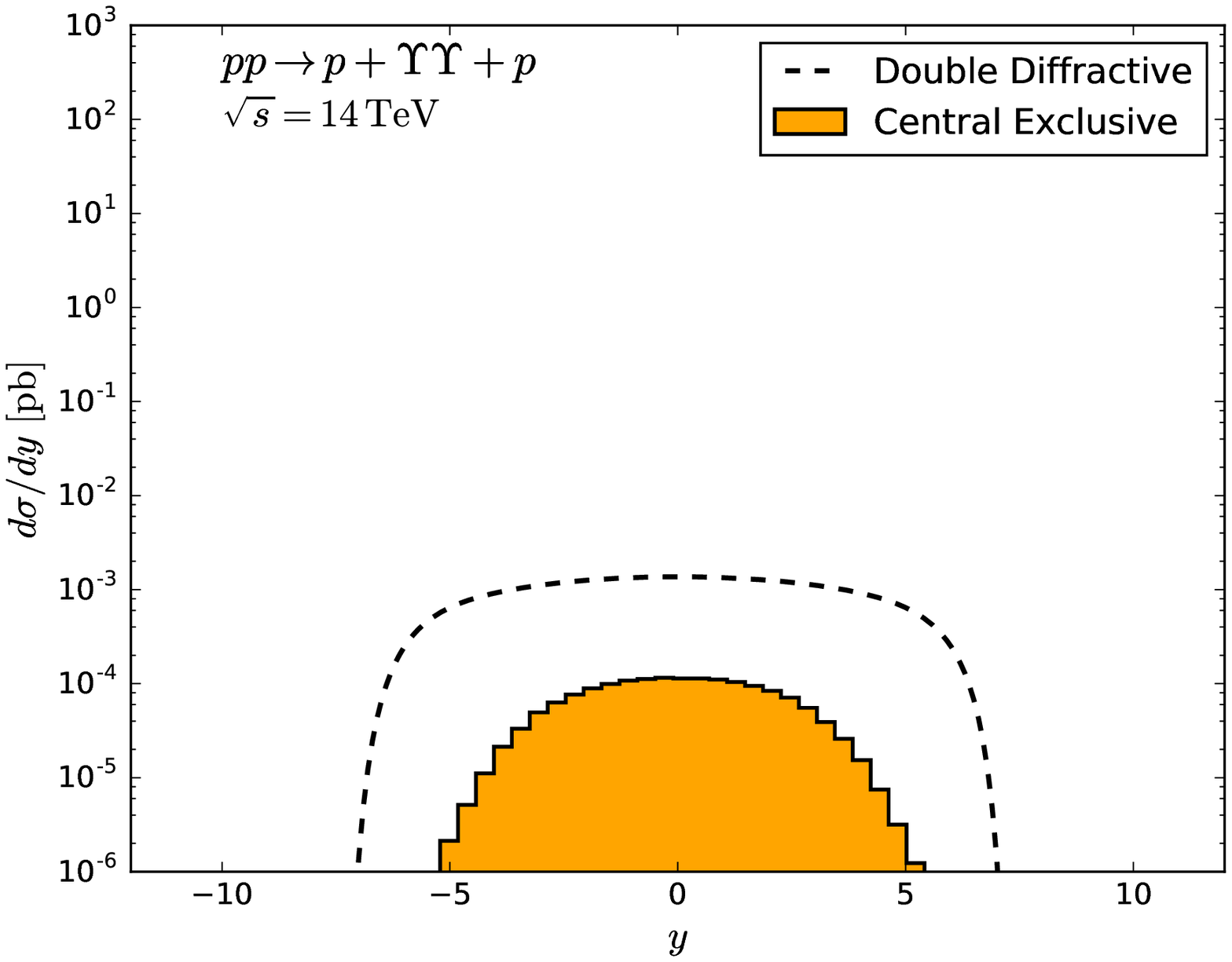}
}
{
\includegraphics[scale = .4]{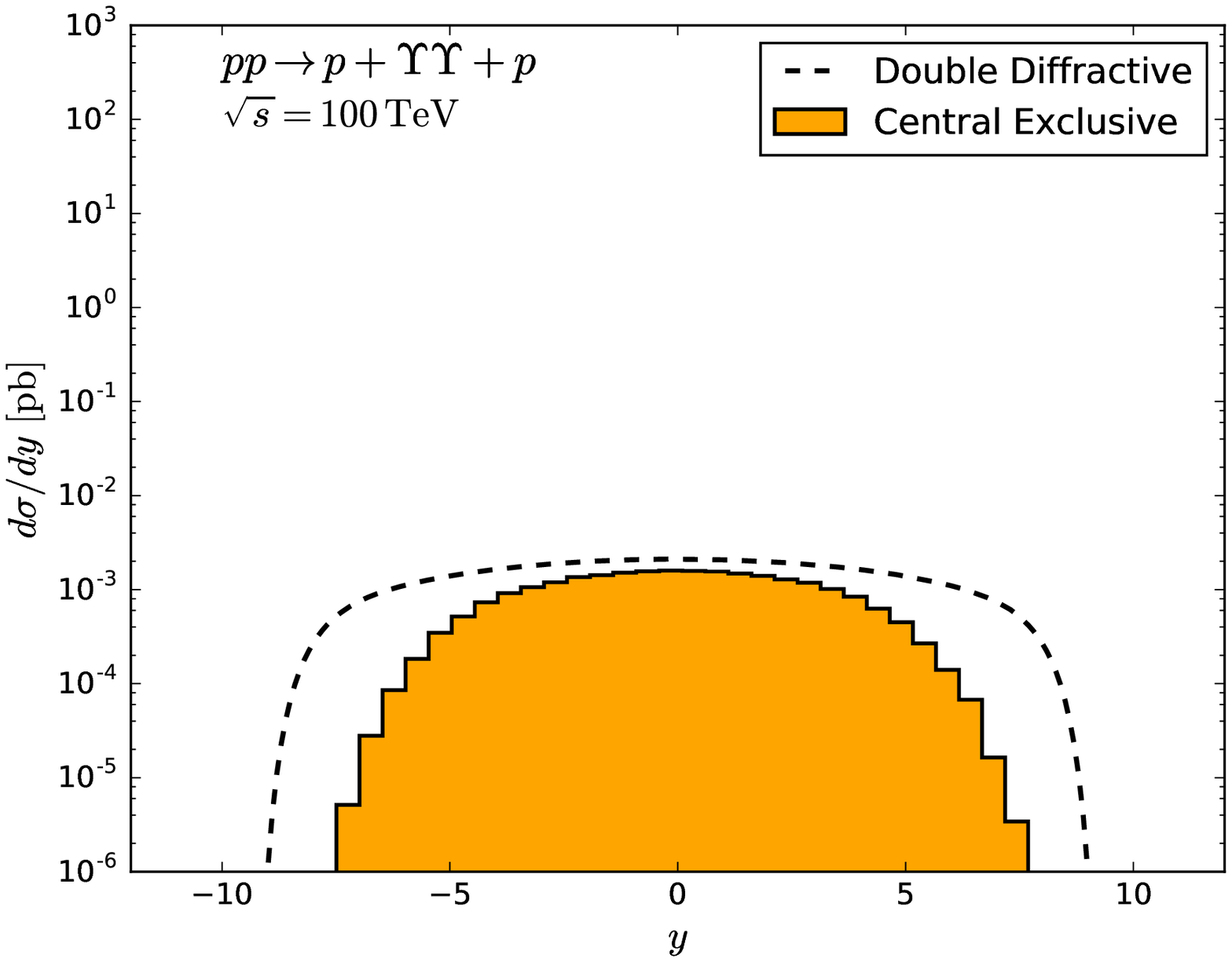}
}

\end{tabular}
\caption{Rapidity distributions for diffractive and central exclusive double $J/\Psi$ (upper panels) and double $\Upsilon$  (lower panels) production in $pp$ collisions at the LHC (left panels) and FCC (right panels) energies.} 
\label{Fig:Rapidez}
\end{figure}

In what follows we will present our predictions for the transverse momentum and rapidity distributions as well for the cross sections considering the diffractive and central exclusive quarkonium -- pair production in $pp$ collisions at the LHC and FCC energies. We will present results for the typical rapidity ranges covered by central ($ -2.5 \le y \le 2.5$) and forward ($ 2.0 \le y \le 4.5$) detectors. For the cross sections we also will present our predictions for $\sqrt{s} = 27$ TeV, which is the center -- of -- mass energy expected to be achieved in the High -- Energy Large Hadron Collider (HE-LHC) \cite{he-lhc}.
In our analysis we will assume that 
$|R_{J/\Psi}(0)|^2=0.56\,$ GeV$^3$ and $|R_{\Upsilon}(0)|^2=2.21 \,$ GeV$^3$.
The predictions for the transverse momentum distribution are presented in Fig. \ref{Fig:TransMom} considering $pp$ collisions at $\sqrt{s} = 14$ TeV. We have verified that similar results are obtained for the energies of HE-LHC and FCC, with the main difference being the normalization of the distributions.
For the double diffractive production, we have that the distribution decreases with $p_{\perp}$ following a power - law behavior $\propto 1/p_{\perp}^n$, where the effective power $n$ is dependent of the final state considered. Such behaviour is expected, since the quarkonium -- pair in the final state in diffractive interactions is generated in a  $2 \rightarrow 2$ subprocess.  In contrast, in the exclusive production, we have that the typical transverse momentum of the quarkonium -- pair is determined by the transferred momentum  in the Pomeron - proton vertex. As the exclusive cross section has an $e^{ - \beta |t|}$ behavior, where $\beta$ is the slope parameter associated,  the associated $p_{\perp}$ distribution decreases exponentially at large transverse momentum. Therefore, it is expected that the production of a quarkonium -- pair  with a large $p_{\perp}$ should be dominated by the  diffractive mechanism.
On the other hand, if only events with $p_{\perp} \le 1$ GeV are selected, the  observed quarkonium - pairs will be mainly produced by the exclusive process. It is important to emphasize that our results also indicate that the contribution of the diffractive process for the double $\Upsilon$ production will not be negligible at small -- $p_{\perp}$.

In Fig. \ref{Fig:Rapidez} we present our predictions for the rapidity distributions considering $pp$ collisions at LHC (left panels) and FCC (right panels) energies. We have that the diffractive mechanism implies wider distributions. Moreover, our results indicate that the production of a double $J/\Psi$ at midrapidities will be dominated by the central exclusive process, with the dominance increasing with the energy. In contrast, we predict the dominance of the diffractive process in the case of double $\Upsilon$ production at the LHC. For the FCC energy, our results indicate that the contribution of the diffractive and central exclusive mechanisms will be similar. 

In Table \ref{Tab2} we present our predictions for the cross sections  considering $pp$ collisions for the center -- of -- mass energies of the LHC, HE -- LHC and FCC, and different rapidity ranges. For the HE -- LHC energy we assume that 
$\langle S^2\rangle$ = 0.015. We predict cross sections of order of pb (fb) in the case of the double $J/\Psi$ ($\Upsilon$) production, which increase with the energy  and are smaller in the forward rapidity range. For the central exclusive processes the increasing is steeper, which is expected since the cross section is proportional to the forth power of the conventional gluon distribution while  in the DD case the cross section is proportional to the square of the diffractive gluon distribution. In agreement with the results presented in Fig. \ref{Fig:Rapidez}, we have that double $J/\Psi$ production is dominated by the central exclusive production. On the other hand, for the double $\Upsilon$ production at the LHC, the DD process dominates. For larger energies, the contribution of the double diffractive and centra exclusive processes becomes similar.

\begin{table}[t]
	\centering
	\begin{tabular}{|| l| l |r | r | r ||}
		\hline
		Energy & Process & Full rapidity range & $ -2.5 \le y \le 2.5$ & $ 2.0 \le y \le 4.5$ \\
		\hline
		\hline
		14 TeV & $\sigma_{DD}(pp\to p+J/\Psi J/\Psi+p)$ & 10.2 pb & 3.7 pb & 1.7 pb \\
		~ & $\sigma_{CEP}(pp\to p+J/\Psi J/\Psi+p)$ & 39.3 pb & 25.9 pb & 5.1 pb \\
		~ & $\sigma_{DD}(pp\to p+ \Upsilon \Upsilon+p)$ & $1.2\times 10^{-2}$ pb & $6.5\times 10^{-3}$ pb & $2.7\times 10^{-3}$ pb \\
		~ & $\sigma_{CEP}(pp\to p+ \Upsilon \Upsilon+p)$ & $1.6\times 10^{-3}$ pb & $1.3\times 10^{-3}$ pb & $2.7\times 10^{-5}$ pb \\
		\hline
		27 TeV & $\sigma_{DD}(pp\to p+ J/\Psi J/\Psi+p)$ & 10.3 pb & 3.9 pb & 1.8 pb \\
		~ & $\sigma_{CEP}(pp\to p+ J/\Psi J/\Psi+p)$ & 85.0 pb & 50.9 pb & 18.3 pb \\
		~ & $\sigma_{DD}(pp\to p+ \Upsilon \Upsilon+p)$ & $1.5\times 10^{-2}$ pb & $7.3\times 10^{-3}$ pb & $3.0\times 10^{-3}$ pb \\
		~ & $\sigma_{CEP}(pp\to p+ \Upsilon \Upsilon+p)$ & $3.9\times 10^{-3}$ pb & $2.8\times 10^{-3}$ pb & $7.5\times 10^{-4}$ pb \\
		\hline
		100 TeV & $\sigma_{DD}(pp\to p+J/\Psi J/\Psi+p)$ & 11.4 pb & 4.0 pb & 2.0 pb \\
		~ & $\sigma_{CEP}(pp\to p+ J/\Psi J/\Psi+p)$ & 439.2 pb & 222.3 pb & 90.4 pb \\
		~ & $\sigma_{DD}(pp\to p+\Upsilon \Upsilon+p)$ & $4.7\times 10^{-2}$ pb & $2.0\times 10^{-2}$ pb & $8.8\times 10^{-3}$ pb \\
		~ & $\sigma_{CEP}(pp\to p+\Upsilon \Upsilon+p)$ & $2.4\times 10^{-2}$ pb & $1.4\times 10^{-2}$ pb & $5.1\times 10^{-3}$ pb \\
		\hline
	\end{tabular}
	\caption{Cross sections for the quarkonium -- pair production in  double diffractive (DD) and central exclusive processes (CEP) considering $pp$ collisions at the LHC, HE-LHC and FCC energies.}
	\label{Tab2}
\end{table}

Finally, let us summarize our main conclusions. In this paper we have investigated the quarkonium -- pair production in double diffractive and central exclusive processes considering  $pp$ collisions at the LHC, HE -- LHC and FCC energies. For the treatment of the double diffractive production we have used the 
 nonrelativistic QCD (NRQCD) factorization formalism for the quarkonium production and the Resolved Pomeron model to describe the diffractive processes. On the other hand, in the case of central exclusive processes, we have considered the Durham model to describe the interaction. We estimated the rapidity and transverse momentum dependencies of the cross sections for the $J/\Psi J/\Psi$ and $\Upsilon \Upsilon$ production and presented  predictions considering the  kinematical rapidity ranges probed by central and forward detectors. The  absorptive corrections have been included in our calculations assuming a simplistic model to treat the soft rescattering corrections. Our results demonstrated that the contribution of the central exclusive (double  diffractive)  processes can be separated selecting events where the transverse momentum of the pair is small (large).  Our results indicate that the study of the quarkonium -- pair production can be useful  to test the underlying assumptions present in the description of the double diffractive and central exclusive processes.

\begin{acknowledgments}
VPG acknowledge useful discussions with  M. Rangel and R. McNulty. This work was  partially financed by the Brazilian funding agencies CNPq, CAPES,  FAPERGS and  INCT-FNA (process number 464898/2014-5).
\end{acknowledgments}

\hspace{1.0cm}

\end{document}